%% file: main.tex
\documentclass{article} % For LaTeX2e
\usepackage{colm2024_conference}

\usepackage{microtype}
\usepackage{hyperref}
\usepackage{url}
\usepackage{booktabs}
\usepackage{graphicx}
\usepackage{listings}
\usepackage{comment}
\usepackage{float}

\colmfinalcopy % Uncomment for camera-ready version, but NOT for submission.

\title{MathViz-E: A Case-study in Domain-Specialized Tool-Using Agents}
\author{Arya Bulusu$^{1}$, 
  Brandon Man$^{2}$, 
  Ashish Jagmohan$^{1}$,
  Aditya Vempaty$^{1}$, \And
  Jennifer Mari-Wyka$^{1}$,
  Deepak Akkil$^{1}$ \\
  \\ $^{1}$ Emergence AI
  \\ $^{2}$ Massachusetts Institute of Technology
}
\date{March 2024}

\begin{document}

\maketitle
\begin{abstract}
There has been significant recent interest in harnessing LLMs to control software systems through multi-step reasoning, planning and tool-usage. While some promising results have been obtained, application to specific domains raises several general issues including the control of specialized domain tools, the lack of existing datasets for training and evaluation, and the non-triviality of automated system evaluation and improvement. In this paper, we present a case-study where we examine these issues in the context of a specific domain. Specifically, we present an automated math visualizer and solver system for mathematical pedagogy. The system orchestrates mathematical solvers and math graphing tools to produce accurate visualizations from simple natural language commands. We describe the creation of specialized data-sets, and also develop an auto-evaluator to easily evaluate the outputs of our system by comparing them to ground-truth expressions. We have open sourced the data-sets and code for the proposed system.  
\end{abstract} 

\input{introduction}

\input{related_work}

\input{methodology}

\input{experiments}
 
\input{discussion}

\subsubsection*{Acknowledgments}
We would like to thank Ravi Kokku, Marc Pickett, Prasenjit Dey, and Paul Haley for helpful discussions and feedback.

\bibliography{bibliography}
\bibliographystyle{colm2024_conference}

\end{document}

%% file: introduction.tex
\section{Introduction}

Large Language Models (LLMs) and Large Multimodal Models (LMMs) have had extraordinary recent success in tasks involving natural language generation and code generation. Spurred by this, there has been significant interest in harnessing LLMs to control software systems and embodied agents through multi-step reasoning, planning and leveraging tools and APIs \cite{karpas2022mrkl, hosseini2021compositional, hao2023toolkengpt, schick2023toolformer, patil2023gorilla}. Promising results have been obtained in many tasks, from device and web-control to game-playing and robotics\cite{wen2024autodroid, lutz2024wilbur, wang2023voyager, wang2023jarvis1, ahn2022i}.

The creation of AI-driven automated systems for specialized domains holds great economic promise; estimated by a recent study at more than a trillion dollars \cite{mckinsey2024agents}. While there exist several general multi-agent frameworks that can be built on, such as Autogen \cite{wu2023autogen, park2023generative}, the development of LLM-driven agentic workflows in specialized domains requires overcoming several additional challenges:
\begin{itemize}
    \item Firstly, such domains come with specialized tools and problems, different from the general-purpose problems that past work has often focused on.
    \item Secondly, there is often a paucity of datasets for training or benchmarking. For example, common math benchmarks like \cite{cobbe2021training}, while important, are of limited value for settings like math pedagogy.
    \item Thirdly, automated evaluation of such systems is hard, and human evaluation does not scale. This makes it hard to create continuous improvement loops, which are essential for robustness.
\end{itemize}

In this paper, we investigate the above issues in the context of a specialized domain; that of math pedagogy. Teachers use a variety of in-classroom technological tools in day-to-day instruction. The variety and complexity of operating these tools imposes a cognitive and time-overload, that teachers would rather spend with students. Generative AI has significant potential to simplify the tools available in the classroom, allowing teachers to spend more time interacting with their students instead of their technology \citep*{mckinsey}. The combination of specialized tool-use, and the potential benefits of automation make classroom pedagogy a well-suited use-case for our exploration.

Reflecting the challenges described above, previous LLM-based math research \cite{mitra2024orcamath, yu2023metamath, liu2023tinygsm, trinh2024alphgeo} has focused on solving math problems and theorem-proving which, while important, are tangential to in-classroom teaching. Existing math benchmarks like GSM8K \cite{cobbe2021training} and MATH \cite{hendrycks2021math} are also of limited value for understanding how LLMs can be applied to the classroom setting. There are no comprehensive datasets that are purposefully aligned with educational standards for middle and high school, nor are there datasets for math visualization pedagogy.

We present an automated math graphing system, that we term MathViz-E, for mathematical pedagogy. Graphs are an essential tool in the classroom, allowing students to visualize and interact with mathematical concepts \citep{donnelly}. Our automated graphing system takes in voice utterances, converts them to mathematical expressions, and graphs them with the Desmos graphing calculator \citep{desmos}. This simplifies the process of creating graphs in the classroom, allowing teachers to more easily incorporate math visualization techniques into their lessons without disrupting classroom flow. 

Our contributions are: 
\begin{itemize}
    \item We present a voice-driven automated graphing system, combining an LLM with a mathematical solver and a visual graphing calculator, for the domain of in-classroom math pedagogy.
    \item We design new domain-specific datasets for graphing problems, representative of the Common Core Math standards \citep{commoncore}, focused on a range of learning objectives that teachers currently use visualization tools to teach.
    \item Evaluating the accuracy of an automated visual graphing system is non-trivial, given that the output of the system is a set of math visualizations. We create an auto-evaluation pipeline to simplify the evaluation of different versions of our system.
    \item We present results demonstrating that our proposed system achieves high accuracy on a wide variety of learning objectives, and show that incorporating multiple tools significantly out-performs an LLM-only system.
\end{itemize} 

The incorporation of multiple tools, including a solver, provides a foundation of accuracy, as LLMs alone are incapable of reliably solving several types of math problems (as we will see in Section \ref{sec:experiments}). This allows the system to produce accurate graphs even for difficult, multi-step problems requiring complex reasoning. On the other hand, while mathematical solvers such as Wolfram Alpha \citep*{wolfram} can provide accurate answers for many categories of problems, they are not capable of understanding all types of natural language. An example of this is anaphora in multi-turn conversations, where a query refers to objects already graphed in the calculator (e.g. "Move the function 5 units horizontally"). By using an LLM to orchestrate across the solver and the visual calculator, we create a robust voice- and dialog-based system. Thus the combination of an LLM with specialized tools produces an automated system with strong potential for domain use. We have open-sourced our code and datasets at \url{https://github.com/EmergenceAI/MathViz-E}.

%% file: related_work.tex
\section{Related Work}

Our work in this paper is related to a large body of recent literature in using LLMs for tool-usage, multi-step reasoning, and plan execution by agents. There is also related work in the use of LLMs for mathematical reasoning, and for pedagogy. In this section, we briefly review this literature. 

There has been considerable work in the last couple of years on using LLMs to orchestrate tools and APIs, motivated by the desire to augment language models' strong text generation capabilities with other, more specialized abilities \cite{karpas2022mrkl}. Systems like NL2API \cite{hosseini2021compositional}, ToolkenGPT \cite{hao2023toolkengpt}, Toolformer \cite{schick2023toolformer} and Gorilla \cite{patil2023gorilla} use language models that learn how to invoke parameterized APIs.  Combining multi-step planning and tool-usage enables the creation of embodied agents that can plan and act in virtual or real environments. Examples of this include agents in virtual game environments \cite{baker2022video, wang2023voyager, wang2023jarvis1} and real-world robotic agents \cite{ahn2022i, bousmalis2023robocat, wu2023unleashing, bhateja2023robotic}. Beyond single-agent systems, there has also been much interest in multi-agent systems and frameworks \cite{wu2023autogen, park2023generative}. In comparison to the above, our work focuses on a narrow set of tools, but for domain-specific capabilities rather than general-purpose usage; we investigate math solvers and visual calculators specifically for mathematical pedagogy. 

There has also been significant work in multi-step reasoning and sequential decision making. Chain-of-thought (CoT) \cite{wei2023chainofthought}, and its many variants \cite{chu2023survey} have shown gains on several types of reasoning tasks. While initially considered an emergent ability in large models like PaLM \cite{chowdhery2022palm}, subsequent work has used techniques like distillation \cite{fu2023specializing, li2023textbooks} to specialize smaller models for specific reasoning tasks.  CoT-style step-by-step reasoning can be further combined with self-critique driven refinement \cite{madaan2023selfrefine, bai2022constitutional}. Another type of approach is to generate code via LLMs, that can be run via tools like Python interpreters, e.g. \cite{gao2023pal}. Also related are approaches like ReAct \cite{yao2023react}, Reflexion \cite{shinn2023reflexion} and many others (e.g. \cite{aksitov2023rest}) that employ LLMs for sequential decision making. In the system we describe in the next section, we use chain-of-thought with a general instruction-tuned LLM for multiple purposes, including query reformulation and tool control. While in this paper we've used vanilla CoT and a large model, the use of more specialized CoT variants and smaller distilled models are intriguing directions for future investigation. 

Also related is the literature on training LLMs for mathematical reasoning. A popular dataset (among many) is the GSM-8K dataset \cite{cobbe2021training} with grade-school math word problems. Recent work has explored the training of small, parameter-efficient models, generally through fine-tuning using augmented data; examples include \cite{mitra2024orcamath, yu2023metamath, liu2023tinygsm}. Finally the use of LLMs for various pedagogical tasks includes work on assessment-generation \cite{wang2022towards, elkins2023useful, bulathwela2023scalable}, learning-content generation~\cite{diwan2023ai, rodway2023impact}, and the use of commercial models like ChatGPT through prompt engineering \cite{adeshola2023opportunities, baidoo2023education}. In contrast to these, we focus on in-classroom math pedagogy, wherein a teacher controls math tools through voice and language.

%% file: methodology.tex
\section{Methodology}
\subsection{Dataset Construction}
The Common Core standards \citep{commoncore} are a set of national educational standards describing what students are expected to know at each grade level, and they have been widely adopted in the United States. Based on the math Common Core standards, we identify a set of learning objectives that teachers use visualization tools to teach in the classroom. We use these categories as the basis of our dataset, creating approximately 10 questions per category to evaluate our system on. 

The categories included in our datasets and the style of utterances were refined through teacher feedback, to reflect language that is commonly used by teachers in math pedagogy (e.g. "Graph a unit circle" rather than "Draw a circle of radius 1".). This feedback also informed the type of problems we chose and the learning objectives we targeted in our datasets. From the identified categories, we create three datasets; the first (referred to as the utterance-focused dataset) is focused on use cases a teacher might want to have available in the classroom. The utterances in this dataset are written as commands a teacher might say, as opposed to written-out problems for a student to solve.  The dataset is mainly comprised of simpler, single-step problems that a teacher might use to demonstrate intermediate steps in the process of solving a problem. To ensure robust evaluation, we created variants for utterances in each category.

\begin{table}[H]
  \caption{Example row from utterance-focused dataset}
  \label{sample-table}
  \centering
  \begin{tabular}{lp{5cm}l}

    \midrule
    Processed Utterance  & Natural Language Utterance    & Graph Input \\
    \midrule
    Reflect y = 5x - 4 across the y-axis & Reflect y equals five x minus four across the y-axis  & $y = -5x - 4  $   \\
    \bottomrule
  \end{tabular}

  \caption{Example question from multi-turn dataset}
  \label{sample-table-multi}
    \begin{tabular}{ll}

    \midrule
    Processed Utterance    & Graph Input \\
    \midrule
    Plot a line that goes through (1, 3) and (4, 8) & $y = \frac{5x}{3} + \frac{4}{3}$    \\ [3pt]
    Plot a parallel line through the origin & $y = \frac{5x}{3}$,  $y = \frac{5x}{3} + \frac{4}{3} $ \\ [3pt]
    \bottomrule
  \end{tabular}  
\end{table}

Our second dataset (referred to as the textbook-focused dataset) is focused on multi-step, complicated problems that require tool use to solve. The main topics in this dataset are a superset of those in the teacher-focused dataset, but the problems are geared towards demonstrating the utility of tool use in LLMs. In contrast to the utterance-focused dataset, we include word problems. The problems in this dataset are based on representative problems explicitly written for Common Core standards.  

Our third dataset, referred to as the multi-turn dataset, is similar to the utterance focused dataset but includes multiple turns in each question. This requires the system to incorporate an understanding of the current calculator state into its response. 

The datasets include a column for the problems and a column for the graph input associated with the problem. As the system is meant to be used through natural language commands, we also included a column with the utterance for the problem (e.g. “Graph $y = 5x^2 + 3$” vs. “Graph y equals five x squared plus three”). This column was automatically generated with GPT-4 \citep{openai2023gpt4} based on the original problem column and manually checked over for accuracy. The utterance-focused dataset contains 70 queries across seven categories, the textbook-focused dataset contains 147 queries across fourteen categories, and the multi-turn dataset contains 95 utterances across seven categories. 

\subsection{System}

The MathViz-E system consists of four main components: the creation of the solver query, the generation of a written explanation based on the solver’s output, the generation of the visual calculator graphing expressions based on the solver’s output, and the validation and correction of the graphing expressions based on LLM self-critique. We incorporate multi-turn functionality into the system by including a calculator state in our prompts which describes the current expressions graphed on the calculator. As a result, the system is able to understand queries within the context of the expressions that have already been graphed. In this paper, we use Desmos as visual calculator, Wolfram Alpha as the solver and GPT-4 as the LLM, but the main principles of the system can be broadly applied to other tools and LLMs. The system is voice-driven; the presented version uses the Web Speech API \cite{mdn-api} for speech recognition, but other ASR pipelines can also be used.

For a given problem, we create the solver query by prompting the LLM with instructions and a series of examples demonstrating how to write queries for certain math problems. These examples were chosen by identifying problems the LLM consistently misunderstood. The LLM is also provided with the spoken-utterance version of the problem and the calculator state. The calculator state contains the equations that have previously been graphed in the graphing window, and passing this state allows the system to incorporate this information into its problem-solving process. Below is a truncated version of the prompt used to create the Wolfram Alpha query:

\begin{quote} 
Write a Wolfram Alpha query that can be used to solve the problem. The main purpose of the task is to find the numerical answer to the problem, not to graph the problem. When writing a query for a word problem, only include the necessary equation to solve the problem. Ensure that the query is acceptable by the Wolfram Alpha engine.  

For example, if you are asked: 

Graph $y = 6x^2 + 4$ and find the local maxima and minima.  

Calculator state: [] 

You generate:  

Find the local maxima and minima of $y = 6x^2 + 4$

\end{quote}

\begin{figure}
  
  \label{fig-system}
  \centering 
  \includegraphics[width=\textwidth]{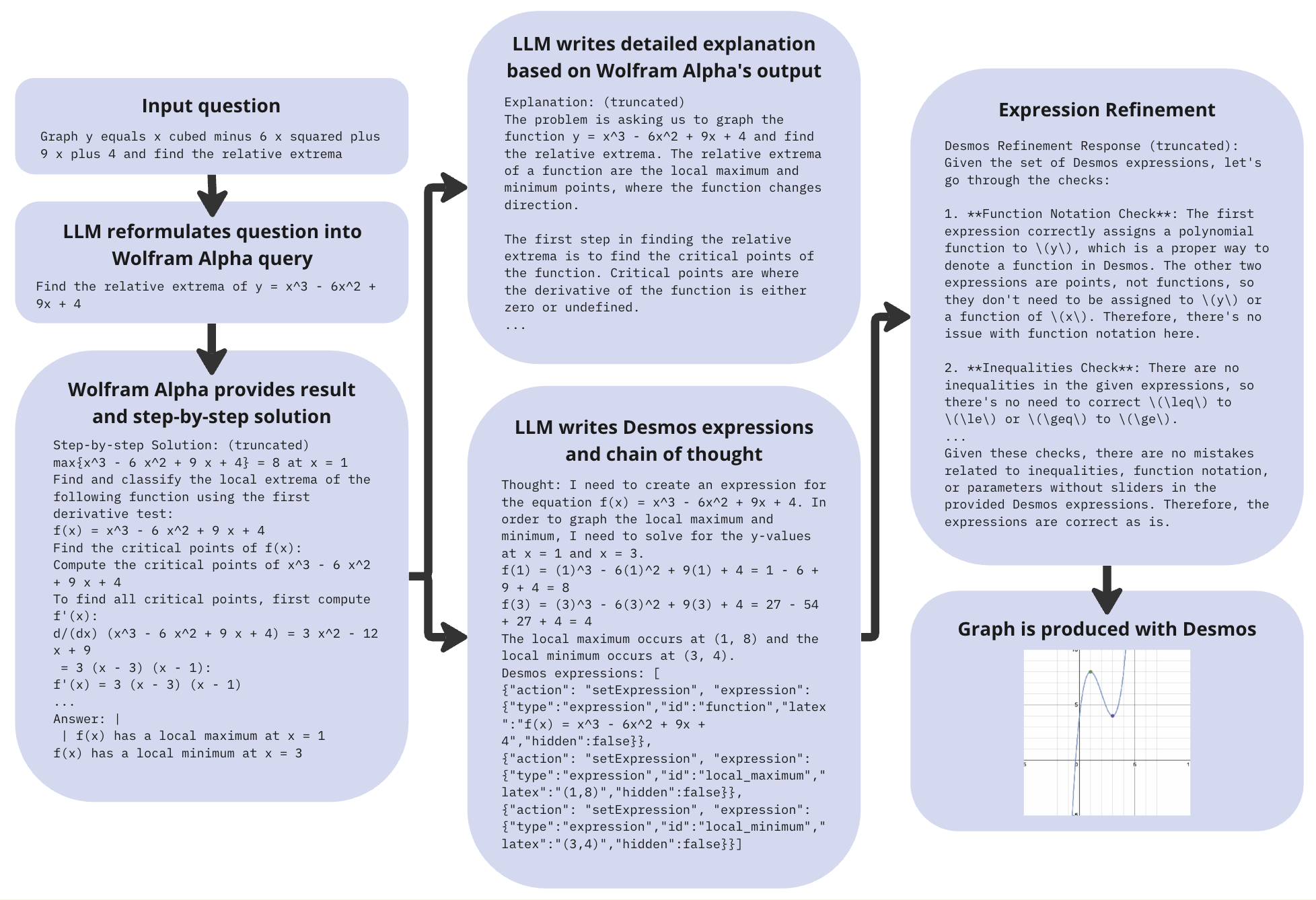}
  \caption{Overview of the MathViz-E automated graphing system}
\end{figure}

	Once the query has been generated, we input it to our solver. Wolfram Alpha provides a set of pods for each query, with each pod containing a different category of information related to the query. Wolfram Alpha also provides step-by-step solutions for some problems. From these results, we extract the solution (generally the second pod, after the “Input Interpretation” pod) and the step-by-step solution if it is present.  
 
To generate an explanation of the problem, we prompt the LLM with a zero-shot instruction. Along with the prompt, we provide the natural language utterance version of the problem, the calculator state, the numerical solution as given by the solver, and the solver’s step-by-step solution, if it is present.

In the cases where Wolfram Alpha provides a step-by-step solution, the LLM only has to expand upon this solution by providing more detail and explaining the reasoning behind the steps. When there is no step-by-step solution given, it must write its own explanation from scratch based on the problem and numerical solution.  

In order to generate the Desmos graphing expressions, we prompt the LLM with instructions and a set of examples. In the prompt, we ask for a chain-of-thought, which helps to generate more accurate expressions. Chain-of-thought prompting has been shown to improve the accuracy of LLMs' reasoning, especially with regards to math problems.\citep{wei}\citep{chu2023survey} As with the explanation prompt, we also provide the natural language utterance version of the problem, the calculator state, the solver’s numerical solution, and the solver’s step-by-step solution, if there is one. The large number of examples in the prompt helps guide the LLM towards producing valid Desmos expressions. The provided examples were created by identifying common points of failure, and writing problems that demonstrate how to accurately deal with these issues.

In the last self-critique step, we ask the LLM to validate and refine the previously-generated Desmos expressions by checking for common errors. A majority of these errors arise from Desmos API deviations from standard latex (e.g. the use of $le$ and $ge$ instead of $leq$ and $geq$ for inequalities, or using $abs(\cdot)$ instead of $|\cdot|$ for the absolute value function). These checks also include ensuring that correct graphing variables are used, operations are formatted correctly, and functions are named properly. This step helps to eliminate basic errors made by the LLM in the expression-generating step.

\begin{figure}
  
  \label{multiturnfigure}
  \centering
  \includegraphics[width=.7\textwidth]{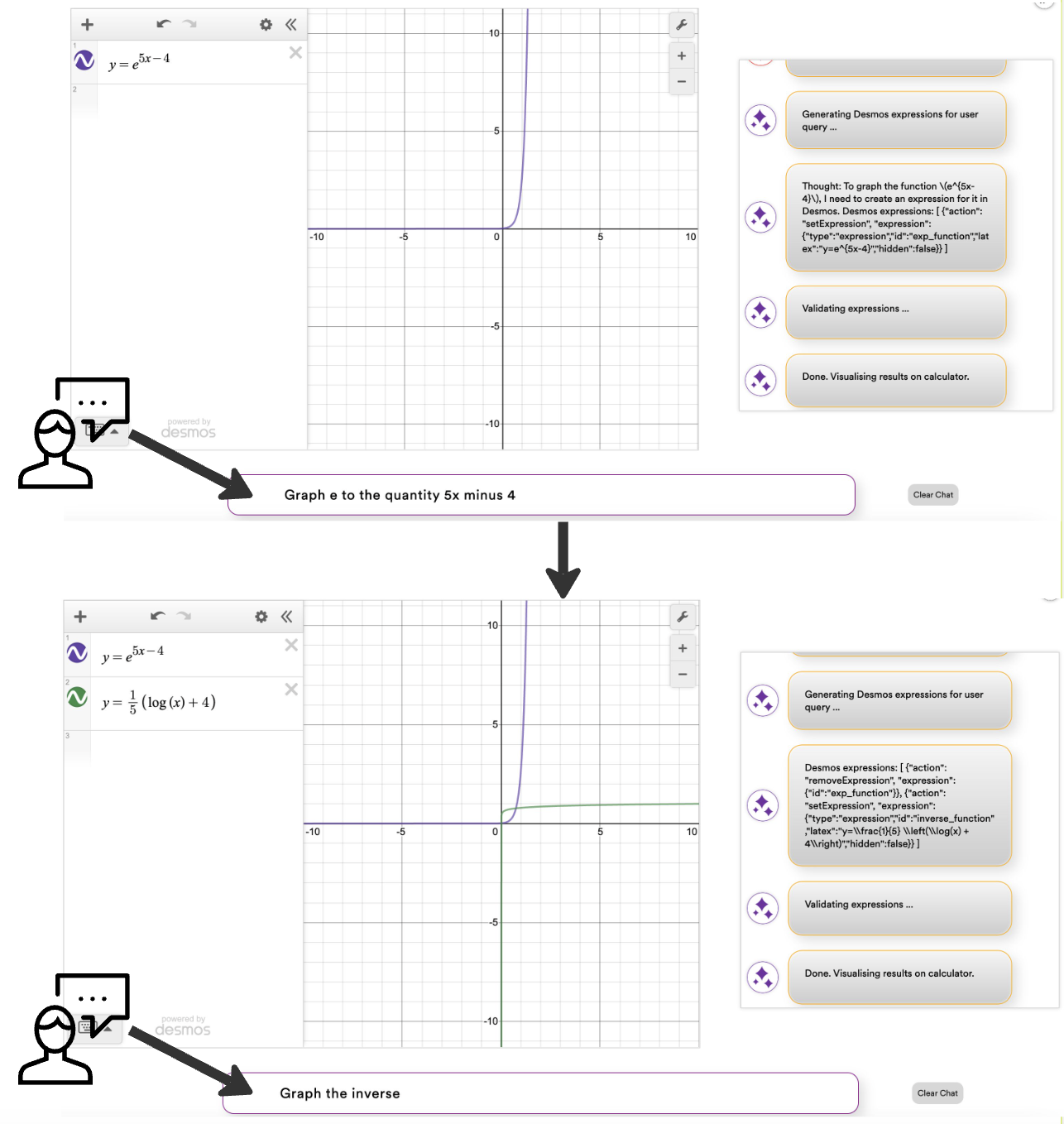}
  \caption{UI of MathViz-E demonstrated through a multi-turn inverse problem}
\end{figure}

%% file: experiments.tex
\section{Experiments} \label{sec:experiments}
\subsection{Autoevaluation}

Traditional evaluation metrics for text similarity fail for comparing mathematical statements due to the precise nature of math statements. Consider the statement 5=2+3. Lexical similarity metrics, such as Jaccard distance, would consider the statement 5=2+4 more similar than 5=4+1 since the former statements shares more words in common than the latter. Furthermore, many existing similarity metrics do not recognize common mathematical symbols as tokens and thus cannot be converted into a numerical representation. Similarly, directly examining the visual graph output through a multimodal approach is unlikely to be precise enough for our purposes. Although LLMs may be able to evaluate equivalence for simple expressions, their judgement becomes inconsistent for more complex expressions. As a result, evaluating the equations output by the automated graphing system at a large scale is nontrivial.

Due to these limiting factors, we create a new autoevaluation pipeline that can precisely compare two mathematical statements. We use the computer algebra system SymPy \citep{sympy} to evaluate the mathematical equations output by the LLM in our LLM+Solver system when responding to given questions. In order to compare two equations, we use SymPy to isolate a variable and compare the resulting expressions on the other side of the equality. 

Although this approach leads to accurate checking of math statements, an issue is that SymPy cannot parse certain formats of equations, which the LLM in the LLM+Solver system may produce. To combat this, we use an LLM as a backup in the autoevaluation process, where if SymPy cannot parse an equation, it will let the LLM compare the two math statements and output a result. 

We construct a set of ground truth evaluations by running all the questions in our datasets through the LLM-only system, and manually evaluating if the system's output matches the correct answer. This manually-benchmarked dataset allows us to run different versions of the autoevaluator on the dataset and check how its evaluations compare to our manually-written evaluations. 

The simpler version of our autoevaluator only uses an LLM to compare two equations. Using GPT-4 as the LLM, we compare the results of LLM-only and LLM+SymPy autoevaluators on the entirety of our utterance-focused and textbook-focused datasets. An older version of the textbook-focused dataset was used, with the same categories and question styles as the current version. In the table below, we display the dataset-wide results as well as the results for selected categories.

\begin{table}[h]
  \caption{Accuracy of LLM-only autoevaluator and LLM+SymPy autoevaluator as compared to manual evaluations}
  \label{autoeval-results}
  \centering
  \begin{tabular}{p{1.9cm}p{1.7cm}p{1.7cm}p{2.7cm}p{2cm}p{1cm}}

    \midrule
      & Utterance-Focused Dataset & Textbook-Focused Dataset & Systems of Linear Equations & Graph Inverse Functions & Graph Lines \\
    \midrule
    LLM-Only & 77\%  & 76\%  & 50\% & 40\% & \textbf{92\%} \\ 
    LLM+SymPy     & \textbf{86\%} & \textbf{88\%}  & \textbf{100\%} & \textbf{80\%} & 85\% \\
    \bottomrule
  \end{tabular}

\end{table}

The addition of SymPy to the autoevaluation pipeline increases the accuracy of evaluations significantly on almost all categories, especially in Systems of Linear Equations. In general, the LLM+SymPy autoevaluator performs better than the LLM-only autoevaluator on problems that are well-structured but computationally difficult. These problems are easily interpretable and can be solved by SymPy, as it can easily handle complex algebraic manipulations. In contrast, an LLM-only approach would struggle to accurately carry out algebra. 

We see a minor drop in performance for a few categories, such as Graph Lines. This decrease in accuracy is generally due to SymPy and the LLM both misunderstanding the formatting of the equation. An important point to note is that the overwhelming majority of incorrectly evaluated answers are the result of a correct input being marked as incorrect. This occurs because SymPy will only mark expressions equivalent if they are genuinely equivalent, and GPT-4 rarely marks inequivalent expressions as being equivalent. As a result, when using the autoevaluation pipeline, we can trust that nearly all the responses marked as correct are actually correct, and manually check if the responses marked as incorrect are in fact incorrect. As a result, this greatly reduces the burden of evaluation when testing the performance of various iterations of the system.

\subsection{Results}

For the results reported in this section, we comprehensively evaluate the LLM+Solver system by manually validating all of the outputs generated for the three datasets (utterance-focused, textbook-focused, and multi-turn). We compare the performance of the LLM+Solver system to the results of the LLM-only system. The LLM-only system consists of directly prompting an LLM with instructions to write Desmos expressions and examples, while also providing the natural language utterance problem and the calculator state. No solver solution is provided. Although the framework of the system can be applied to LLMs and solvers broadly, in this paper we evaluate using GPT-4 as the LLM and Wolfram Alpha as the solver. Tables \ref{table-results-1}, \ref{table-results-2} and \ref{table-results-3} compare the results for the LLM-only system and the LLM+Solver systems.

\begin{table}[th]
  \caption{Accuracy of LLM-only and LLM+Solver models \\}
  \label{table-results-1}
  \centering
  \begin{tabular}{lll}

    \midrule
         & LLM-only     & LLM+Solver \\
    \midrule
    Utterance-Focused Dataset & 66\%  & \textbf{90\%}     \\
    Textbook-Focused Dataset     & 64\% & \textbf{86\%}      \\
    Multi-turn Dataset     & 86\% & \textbf{91\%}      \\
    \bottomrule
  \end{tabular}
\end{table}

\begin{table}[th]
  \caption{Accuracy of individual categories in utterance-focused dataset \\}
  \label{table-results-2}
  \centering
  \begin{tabular}{lll}

    \midrule
         & LLM-only     & LLM+Solver \\
    \midrule
    Graph Circles & 90\%  & \textbf{100\%}     \\
    Transform Shapes     & 70\% & 70\%      \\
    Intersections of Lines     & 20\% & \textbf{100\%}      \\
    Transform Functions     & \textbf{100\%} & 90\%      \\
    Graph Lines     & 90\% & \textbf{100\%}      \\
    Local Minima and Maxima     & 40\% & \textbf{90\%}      \\
    X Intercepts, Y Intercepts     & 50\% & \textbf{80\%}      \\
    \bottomrule
  \end{tabular}
\end{table}

\begin{table}[t]
  \caption{Accuracy of individual categories in textbook-focused and multi-turn datasets \\}
  \label{table-results-3}
  \centering
  \begin{tabular}{lll}

    \midrule
         & LLM-only     & LLM+Solver \\
    \midrule
    Proportional Relationships & 100\%  & 100\%     \\
    Linear Inequality Systems   & \textbf{100\%} & 90\%      \\
    Graph Inequalities     & 90\% & \textbf{100\%}      \\
    Graph Lines     & 100\% & 100\%      \\
    Graph Lines (multi-turn)    & 91\% & \textbf{100\%}      \\
    Graph Polynomials + Identify Zeros     & 67\% & \textbf{100\%}      \\
    Graph Polynomials (multi-turn)     & 67\% & \textbf{100\%}      \\
    Systems of Linear + Quadratic Equations     & 15\% & \textbf{100\%}      \\
    Systems of Lin + Quad Eqns (multi-turn)     & 60\% & \textbf{100\%}      \\
    Graph Circles     & 100\% & 100\%      \\
    Graph Circles (multi-turn)    & \textbf{96\%} & 83\%      \\
    Transformations of Functions     & 100\% & 100\%      \\
    Transform Functions (multi-turn)    & 86\% & \textbf{90\%}      \\
    Graph Inverse Functions     & 70\% & \textbf{100\%}      \\
    Graph Inverse Functions (multi-turn)    & 90\% & \textbf{100\%}      \\
    
    Tangents to Parabolas     & 0\% & \textbf{11\%}      \\
    Tangents to Circles     & 0\% & \textbf{75\%}      \\
    Linear and Nonlinear Functions     & 60\% & \textbf{70\%}      \\
    Linear and Nonlinear Functions (multi)    & \textbf{100\%} & 70\%      \\
    Systems of Linear Equations     & 40\% & \textbf{100\%}      \\
    Rigid Transformations + Dilations     & \textbf{70\%} & 60\%
    \\
    
    \bottomrule
  \end{tabular}
\end{table} 

\textbf{LLM+Solver vs. LLM-only system} Table \ref{table-results-1} shows that the addition of the solver results in a significant overall performance increase across all three datasets. Tables \ref{table-results-2} and \ref{table-results-3} further show that the greatest performance increase occurs in categories such as Local Minima and Maxima, X and Y intercepts, Intersections of Lines, Systems of Equations, and Tangents to Circles. These problems require complex calculations which are difficult for GPT-4 to carry out by itself, meaning that GPT-4 will frequently get them wrong. However, Wolfram Alpha can solve these problems, given a well-formulated query. As a result, the LLM+Solver system shows strong improvement over the LLM-only system for these categories.

\textbf{Error Modes for LLM+Solver system} The proposed system demonstrates excellent accuracy across a wide variety of problem types, as demonstrated in Table \ref{table-results-2} and Table \ref{table-results-3}. However, there are some notable categories on which it has poor accuracy. It is instructive to delve deeper into these error modes. \begin{itemize}
    \item For the category "Tangents to Parabolas", the problem arises from the Wolfram ALpha solver API, which does not return the correct answer in its Step-by-Step Solution response. This is a specific instance of a broader problem, where it is not always clear which of the Wolfram Alpha API response fields should be input to the LLM for further reasoning; fixing this requires conditioning response interpretation on the problem type (e.g. "Tangents to Parabola" and "Tangents to Circles" produce different types of API responses). 
    \item For "Rigid Function Transformations", errors arise from improper handling of polygon transformations. While the solution is very robust to transformations of parametric shapes, it struggles with non-parametric shape transformations. Improving this is an area of future work.
    \item For other categories, some common (though occasional) error modes include: incomplete specifications (like "Draw a circle" without any radius or center specified) sometimes result in the system producing parametric expressions without associated sliders; a single query containing a list of sub-queries ("Plot the x-intercept, the y-intercept, the local minima and maxima, and the asymptotes") occasionally result in some sub-queries being missed; and tasks of sufficient complexity ("Move the circle so it is tangent to a line which satisfies property x and property y") may be computed incorrectly. The last error mode is not generally an issue for the pedagogical levels that the system is designed for, but could be an issue for higher-ed (compared to K-12). 
\end{itemize}

\textbf{Analysis of the LLM-only system}  The LLM-only system struggles the most in problems that require complicated reasoning and calculations to solve, such as "Tangents to Circles", "Systems of Linear + Quadratic Equations", "Intercepts" and others. It tends to fail either by solving the problem with an incorrect method or executing calculations incorrectly. It also sometimes plots the underlying functions or shapes, but then does not successfully compute points of interesections, extrema etc. 

For many categories, the performance of the LLM-only system was strong to begin with, such as Circles, Proportional Relationships, and Graph Lines. These categories contain simple problems with little mathematical calculation, so GPT-4 is able to succeed at these problems without the help of a solver. There are some categories for which there are no Wolfram Alpha queries that can be used to solve the problem, such as Transform Shapes and Transformations on Functions. These categories do not show much change in the performance of both systems, as Wolfram Alpha cannot be used to provide answers for these categories. Incorporation of a more powerful tool, such as Python, could allow the system to successfully solve these problems. 

%% file: discussion.tex
\section{Discussion}
The results presented in this paper highlight the potential of domain-specific automation created via LLM orchestration of specialized tools. The presented case-study highlights some of the main challenges that need to be overcome in such systems. The paucity of preexisting datasets requires careful creation of new datasets for benchmarking; there is a need to identify and control specialized tools; and there is a need for auto-evaluation to validate and improve system performance. 

For the case of mathematical pedagogy, we showed that through the design and development of an LLM-orchestrated system incorporating multiple specialized tools, through the creation of new benchmark datasets based on Common Core, and through the creation of an auto-evaluation pipeline, we created an effective automated system and the means to easily evaluate future iterations. The system has strong performance, and for many categories of problems it can consistently produce accurate outputs.  

For the domain under consideration, there are categories of problems for which the system is not yet fully reliable. To improve system performance in these categories, there are several directions that we can take in the future. An especially promising approach is to add retrieval-based techniques to make adjustments to the system for the individual problem categories. This individualized approach could improve accuracy for some categories as compared to our current, one-size-fits-all system. Another key direction is to reduce latency;  we plan to evaluate our system using smaller open-source LLMs, trained via supervised fine-tuning, instead of a large general model like GPT-4.

More broadly, for domain-specific agentic solutions, the approach presented in this paper offers a set of patterns that can be generalized. We expect that the generalization and application of these patterns to other domains and problems will continue to remain fertile ground for future research.